\documentclass[]{aastex631}

\usepackage{graphicx}    % Include figure files
\usepackage{amsmath}

\shorttitle{SNR 0509-67.5 reverse-shocked ejecta}
\shortauthors{Das et al.}

\begin{document}

\title{Ejecta clumps revealed by study of reverse-shocked ejecta through MUSE integral field spectroscopy of SNR 0509-67.5}

\author[0000-0002-5483-0232]{Priyam Das}
\altaffiliation{E-mail: priyam.das@unsw.edu.au}
\affiliation{School of Science, The University of New South Wales, Northcott Dr., Canberra, ACT 2612, Australia}
\affiliation{Research School of Astronomy and Astrophysics, Australian National University, Canberra ACT 2611, Australia}

\author{Ivo R. Seitenzahl}
\affiliation{Research School of Astronomy and Astrophysics, Australian National University, Canberra ACT 2611, Australia}
\affiliation{Mathematical Sciences Institute, Australian National University, Canberra ACT 2611, Australia}

\author{J. Martin Laming}
\affiliation{Space Science Division, Naval Research Laboratory, Washington DC 20375, USA}

\author{Gilles Ferrand}
\affiliation{The University of Manitoba, Department of Physics and Astronomy, Winnipeg, Manitoba, R3T 2N2, Canada}
\affiliation{RIKEN Center for Interdisciplinary Theoretical and Mathematical Sciences (iTHEMS), Wak\={o}, Saitama 351-0198 Japan}

\author{Simon J. Murphy}
\affiliation{School of Science, The University of New South Wales, Northcott Dr., Canberra, ACT 2612, Australia}

\author{Ashley Ruiter}
\affiliation{Mathematical Sciences Institute, Australian National University, Canberra ACT 2611, Australia}

\begin{abstract}
We report the discovery of a spatially resolved clumpy ejecta structure in the reverse-shocked ejecta of SNR 0509–67.5, revealed through multiple faint and broad forbidden coronal emission lines in deep MUSE observations.
We also identify two new broad coronal emission lines not reported before in this remnant, [Fe \textsc{xi}] 7894~\AA\ and [Fe \textsc{x}] 6374.5~\AA , which extend the set of previously reported [Fe \textsc{xv}] 7059.59~\AA, [Fe \textsc{xiv}] 5302.86\AA , [Fe \textsc{ix}] 8236.55~\AA , [Ca \textsc{xv}] 5695\AA, and [S \textsc{xii}] 7611.0~\AA. Near-continuous ionisation states of Fe allow us to follow the ionisation progression behind the reverse shock. We use a 1D analytical model to evolve Fe charge states following reverse shock interaction to compare with observations, indicating the need for preshock clumping or over-density in order to reproduce the observed surface brightness of the [Fe\,\textsc{xiv}] line. Additionally, we report the spatially resolved distribution of ejecta clumps and show that reverse-shock interaction drives their compression and fragmentation. We also find a clear trend of decreasing velocity width with increasing Fe ionisation state, from the broadest [Fe\,\textsc{ix}] emission to the narrowest [Fe\,\textsc{xv}], with intermediate-ionisation species ([Fe\,\textsc{x}], [Fe\,\textsc{xi}], [Fe\,\textsc{xiv}]) showing intermediate widths.
Finally, we compare our observations to a dynamically driven double-degenerate double detonation (D6) 3D remnant model at similar Fe and S ionisation states, and conclude that the observed clumps are predominantly due to Rayleigh–Taylor instabilities.
\end{abstract}

\keywords{supernova remnant -- reverse-shocked ejecta -- Type Ia -- spectroscopy}

\section{Introduction}

Type Ia supernovae (SNe Ia) are thermonuclear explosions of carbon-oxygen white dwarfs~\citep{Hoyle1960}. SNe Ia are differentiated from other stellar explosions by identifying the absence of hydrogen and helium lines in their spectra and the presence of a silicon line at 6150~$\mathrm{\AA}$. Astronomers have used them to study the expansion of the Universe and dark energy for decades as they can be used as standardisable candles \citep{Brian1998,Riess1998,Perlmutter1999}. They are a prime site for nucleosynthesis, producing intermediate-mass elements and stable isotopes of the iron group elements (IGE). Despite their importance for galactic and stellar evolution, the progenitor systems and explosion mechanisms of Type Ia supernovae remain poorly understood \citep{Hillebrandt2013} \citep[see][for recent reviews]{Ruiter2025, Liu2023}. Historically,  there are two main competing scenarios: the Chandrasekhar-mass scenario, where a near-Chandrasekhar (hereafter near-Ch) mass ($\sim$1.4 M$_{\odot}$) white dwarf accretes material from a companion until central ignition \citep{Khokhlov1991,Arnett1994,branch1995a}, and the sub-Chandrasekhar mass scenario, where the explosion is triggered via double detonation from either a white dwarf merger or a helium shell detonation \citep{iben1987interacting,livne1995explosions, sim2010a, Fink2010,Woosley2011, pakmor2022a}.

Regardless of the nature of the explosion, a Type Ia supernova explosion unbinds the stellar material of its progenitor WD into the surrounding ambient medium. Around $\sim$$10^{51}~\mathrm{erg}$ of energy is released in the explosion, which initiates a shock wave (forward shock) that propagates radially outward from the center of the explosion. The forward shock gradually decelerates as it interacts with the surrounding medium, producing an inward-propagating (in a Lagrangian sense) reverse shock~\citep{Ardavan1973,Mansfield1974,Truelove1999,Reynolds2017}. In young supernova remnants, where the ionisation timescale is comparable to the supernova remnant age, the reverse shock heats the ejecta as it passes through and the ejecta progressively ionises with time behind the shock front due to non-equilibrium conditions. Since the supernova remnant does not exist in ionisation equilibrium, the behaviour seen in radiative shocks, where ionisation states decrease over time due to recombination \citep{Itoh1977,Shull1982}, is reversed. Instead, the ionisation of different elemental species increases with time, creating zones of higher ionisation near the outer boundary (old-shocked ejecta) of the ejecta and lower ionisation inwards (newly-shocked ejecta)  \citep{Lu2015, Laming2014}. This highly-ionised reverse-shocked ejecta produces X-ray emission lines due to collisional excitation \citep{Badenes2003,Badenes2006,Ballet2006, Patnaude2012} and forbidden coronal lines in the optical discovered by \citet{Seitenzahl2019} (hereafter IS19). In this paper, we analyse the spatial morphology of the clumps, line profiles and velocity width trends of these forbidden optical coronal lines.
The evolution of the remnant and the interaction of the ejecta with the reverse shock gives rise to hydrodynamical instabilities, which can be detected through the observed morphology of the reverse-shocked ejecta.

The expansion of the dense reverse-shocked ejecta gas is decelerated by the less dense forward-shocked ambient medium. The interaction between the shocked ejecta and the surrounding medium establishes a contact discontinuity, where deceleration drives the growth of finger-like Rayleigh–Taylor (RT) instabilities that penetrate and fragment the  less-dense gas~\citep{rayleigh1903scientific,taylor1950instability}. As the instability evolves, the linear growth phase of the initial perturbation transitions into non-linear growth, forming Kelvin-Helmholtz (KH) caps \citep{Wang2001}. The difference in relative velocities between the gases forms mushroom caps at the ends of the RT instabilities, which facilitates efficient mixing of the gases.  The development of RT instabilities is crucial for the morphology and evolution of the supernova remnants and can be indicative of a particular explosion mechanism \citep{Ferrand2019, Ferrand2021}. Formation of clumps and the irregular morphology of ejecta has been attributed to strong inter-shock convection that transmits the material beyond the forward shock's nominal radius.
%However, the extensively mixed structure, including the dispersal of Ni$^{56}$ to high velocities, is now thought to be induced by the R-T instability via an asymmetric explosion~\cite{Kifonidis2000,Hungerford2003}. 
Another theory that explains the clumping of ejecta material is the nickel bubble effect. The sequential radioactive decay of $^{56}$Ni$\rightarrow$$^{56}$Co$\rightarrow$$^{56}$Fe heats up the ejecta, creating an inflated nickel bubble~\citep{Wang2008, Woosley1988a}. The radiation deposited by radioactivity seeps through the shock via radiative diffusion, influencing the development of the ejecta structure. This creates a density gradient in the ejecta material, increasing radially outwards, with the highest density at the bubble–shell interface \citep{Wang2005}. The passage of the reverse shock through such structured ejecta introduces an impulse, which gives rise to Richtmyer-Meshkov (RM) instabilities \citep{richtmyer1954taylor,Meshkov} in the initial evolution phase of the remnant. Regardless of how the clumps are formed, they are fragmented due to enhanced mixing by RT instabilities at the contact discontinuity \citep{Bykov2022} towards the later stages of their evolution.

\citet{Gull1973a} first modeled supernova remnant evolution and demonstrated RT instability growth in young supernova remnants. Subsequent multi-dimensional studies \citep[e.g.,][]{Chevalier1992a} showed that RT fingers are limited by KH caps and ram pressure, preventing them from reaching the forward shock. Two- and three-dimensional MHD simulations of the RT instabilities in the shell of Type Ia supernova remnants were carried out using a moving grid technique, which allowed for the tracking of the evolution of the RT fingers and their effects on local magnetic fields \citep{Jun1995, Jun1996}. The growth of the RT fingers has been suggested as a cause for the enhancement of magnetic field effects in three dimensions, creating vortices and ring-structured magnetic fields. Although these simulations have significantly advanced our understanding of RT instability growth in supernova remnants, \citet{Wang2011a} pointed out that most simulations use a power-law density distribution for SN Ia ejecta, which is not realistic. Instead, an exponential profile should be followed, as the density of SNe Ia ejecta decreases with expansion and follows a significantly different density profile, which can be well approximated by an exponential function~\citep{Dwarkadas1998, Dwarkadas2000}.

Several observations of different young Type Ia SNRs have revealed clumpy ejecta in X-rays, which are associated with hydrodynamic instabilities~\citep{chevalier1982self,Nadyozhin1985}. High-resolution X-ray imaging of Tycho's supernova remnant with the \textit{Einstein} and ROSAT missions revealed knots and clumps in the ejecta~\citep{Seward1983}. Later, higher resolution  \textit{Chandra} X-ray imaging was able to spatially resolve and determine the separation between the clumps on scales of a few arcseconds~\citep{Hwang2002, Warren2005}. Similar X-ray images from \textit{Chandra} also revealed clumpy ejecta in SNR 1006, along with protrusions of the clumps ahead of the forward shock \citep{Winkler1976,Hamilton1986,Pye1981}. These protrusions were generally interpreted as an effect of back-reaction from cosmic ray acceleration; however, recent MHD studies revealed that energy loss from cosmic ray acceleration alone is insufficient to explain both the separation between the forward shock and the contact discontinuity and the protrusions. Instead, initial clumping or asymmetry might be responsible for the breakouts \citep{Orlando2012,Dyer2014,Sato2019}. \textit{Chandra} observations of SNR 3C 397 and many other SNRs with the Advanced CCD Imaging Spectrometer also reveal arcsecond-scale clumps \citep{Safi-Harb2005, Sato2017a, Sato2017b, Williams2017}.

SNR 0509-67.5 (hereafter SNR 0509) is a young Type Ia supernova remnant located in the Large Magellanic Cloud at a distance of about 50 kpc. It is estimated to be roughly 310–350 years old, based on proper motion analyses of the remnant \citep{arunachalam22}. The mass of the ejecta in SNR 0509 is consistent with a sub-Chandrasekhar mass ($\sim$$1~\mathrm{M}_\odot$), the explosion energy is on the order of $1–1.4 \times 10^{51}$ erg \citep{Hovey2015} and it has been classified as a 91T-like SN from light echo studies \citep{Rest2005}. The remnant is expanding into a low-density ambient medium, with a preshock hydrogen number density of $\sim$$0.05–0.1~\mathrm{cm^{-3}}$ \citep{Ghavamian2017}, producing prominent Balmer-dominated shocks around its periphery. Its nearly perfect circular morphology suggests an isotropic explosion in a relatively uniform environment. Recent observations of a double-shell ejecta morphology of [Ca \textsc{xv}] have suggested its origin from a double detonation explosion scenario \citep{Das2025, Mandal2026}.  See \citet{Soker2025} for an alternative explanation. \citet{Seitenzahl2019} demonstrated radial separation between some lines from the reverse-shocked ejecta of SNR 0509. Motivated by these characteristics, and the growing evidence for structured ejecta in SNR 0509, we present a high spatial resolution optical study of the remnant, revealing spatially resolved clumps in the ejecta shells of iron and sulphur with high S/N.

\section{Methods}
\subsection*{Data Reduction}
We performed a deep observation of SNR 0509 using the MUSE \citep{bacon2010muse} optical integral field spectrograph, mounted on Unit Telescope 4 of the European Southern Observatory's (ESO) Very Large Telescope at Cerro Paranal. This observation was part of ESO program ID 0104.D-0104(A), led by P.I. Seitenzahl. We collected the data in service mode using the WFM-AO setup over 25 separate nights spanning 24 months: 39 observations, each lasting approximately 2700 seconds, and a single observation of 93.92 seconds (which we excluded due to poor quality), resulting in a total exposure time of around 105,300 seconds (29 hours and 15 minutes). Data reduction, night-sky subtraction, and de-reddening procedures are identical to those described in \citet{Das2025} and \citet{Das2026}.

\subsection*{Data Analysis}

\begin{figure*}[h]
    \centering
    \includegraphics[width=20cm, trim={280 40 240 40}, clip]{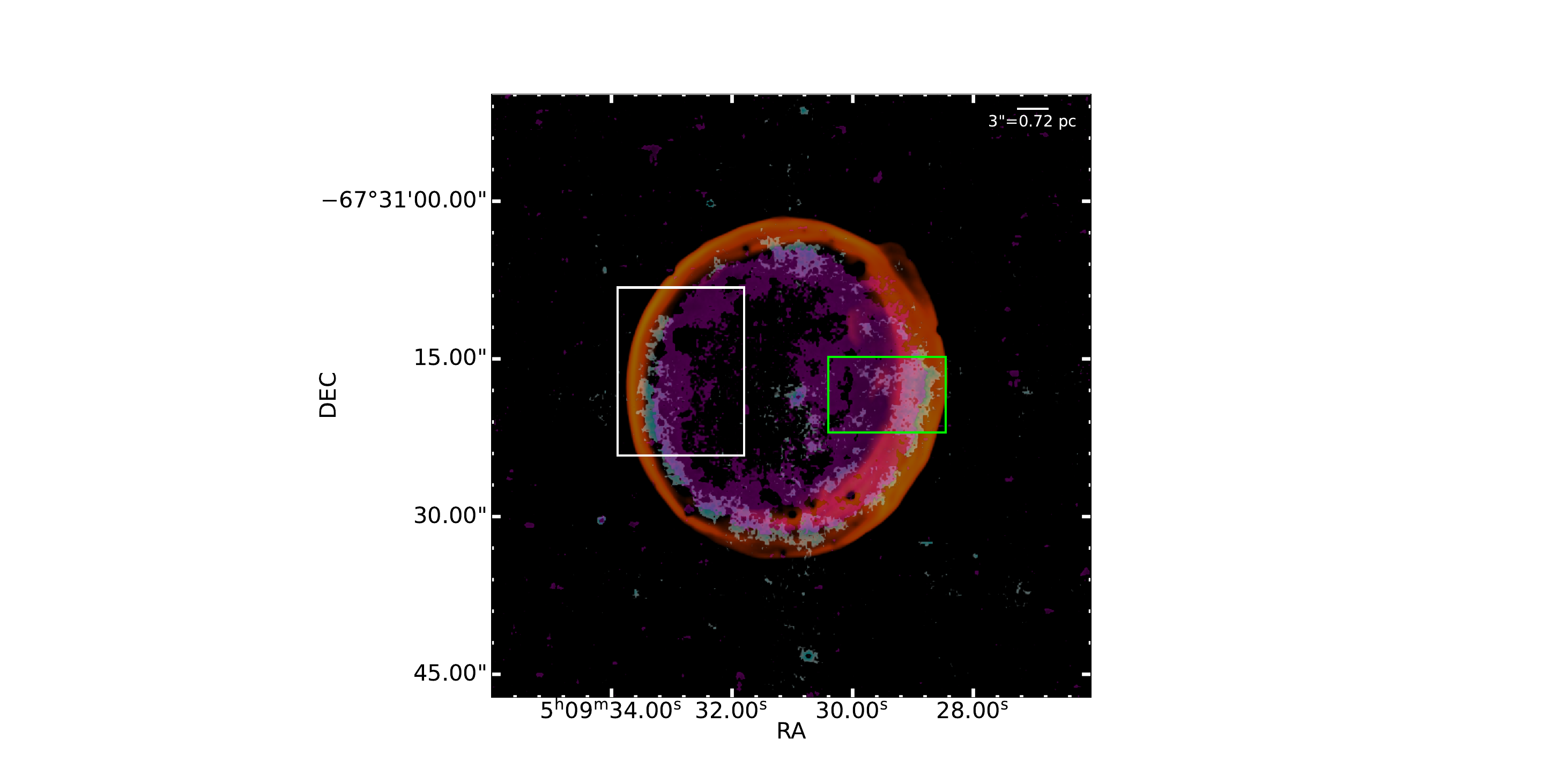}
    \caption{SNR 0509 showing H$\mathrm{\alpha}$ from the Balmer dominated shock region in orange, [Fe \textsc{xiv}] in magenta and [S \textsc{xii}] in cyan from the reverse-shocked ejecta. The area inside the white rectangle is used to analyse the ejecta in the eastern region (see Figs.~\ref{fig:east} and \ref{fig:iron_clumps}), while the area inside the green rectangle is used to analyse the ejecta in the western region (see Figure~\ref{fig:west}); in both regions, line profiles are extracted to derive velocity widths.}
    \label{fig:rgb}
\end{figure*}

\begin{figure*}[h]
    \centering
    \includegraphics[width=18cm, trim={10 0 10 0}, clip]{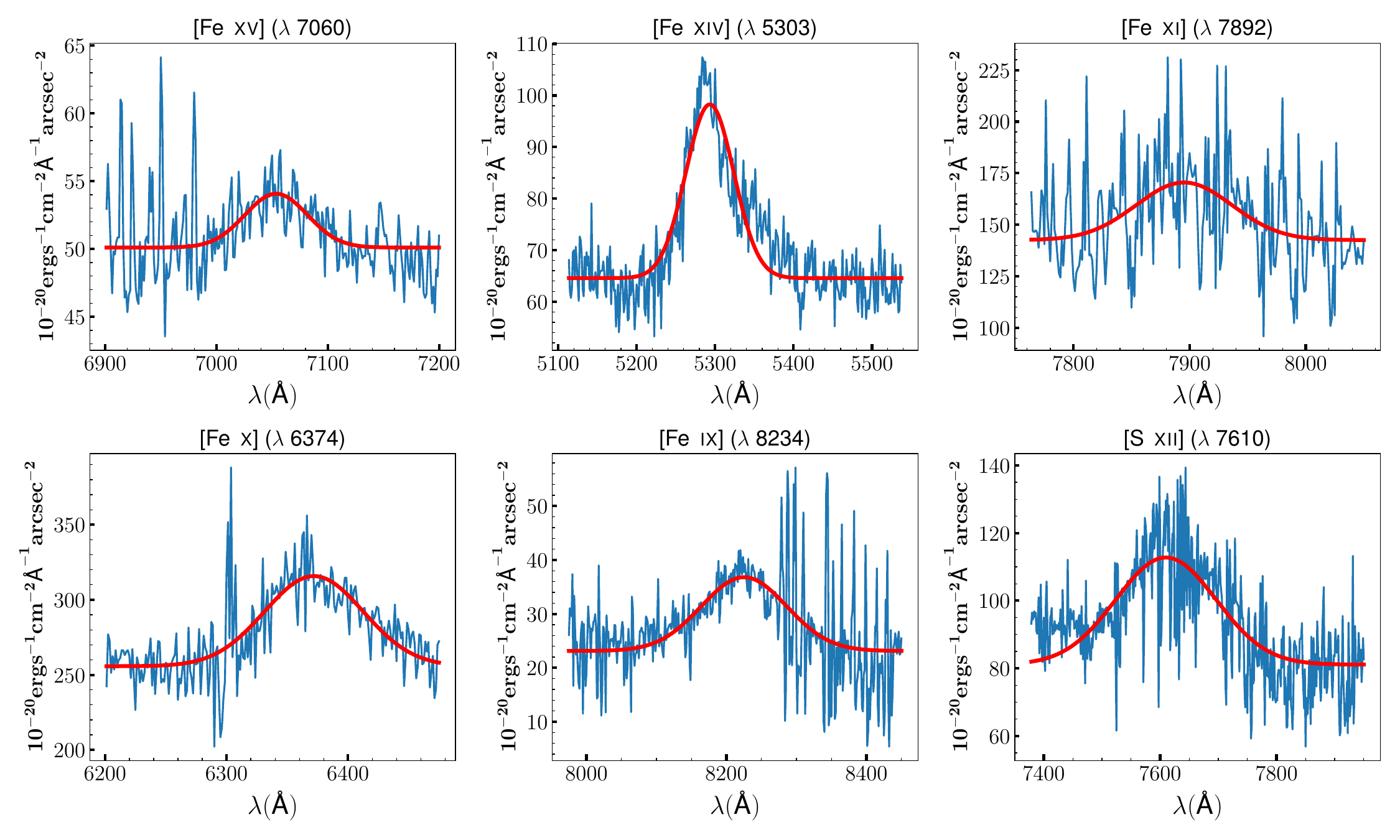}
    \caption{Extracted spectra and single component Gaussian plus linear continuum fits for the broad coronal emission lines of Fe and S from the reverse-shocked ejecta in the eastern region of SNR 0509 (white rectangle in Figure~\ref{fig:rgb}). }
    \label{fig:east}
\end{figure*}

\begin{figure*}[h]
    \centering
    \includegraphics[width=18cm, trim={10 0 10 0}, clip]{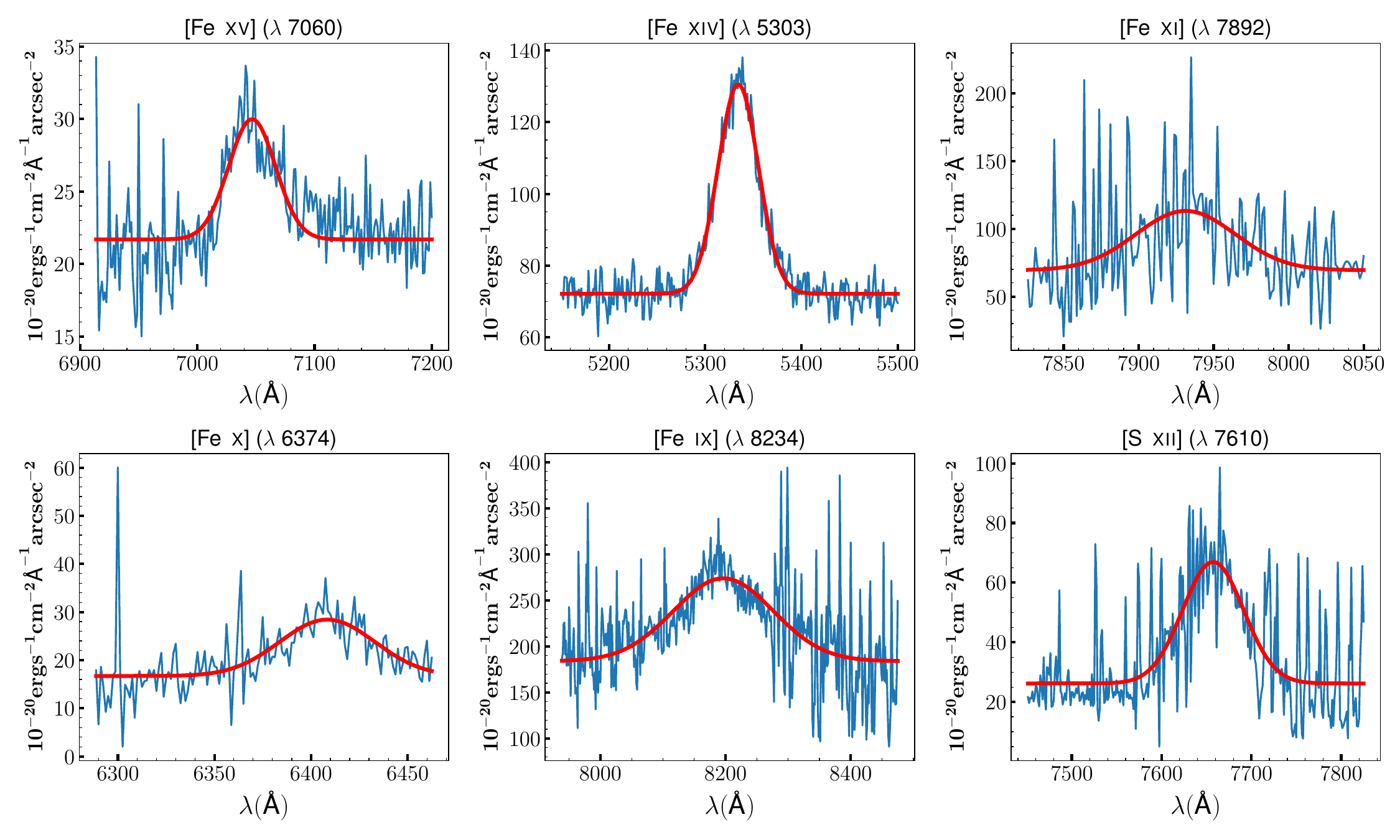}
    \caption{Same as Figure~\ref{fig:east}, except for the western region of SNR 0509 (green rectangle in Figure~\ref{fig:rgb}).}
    \label{fig:west}
\end{figure*}

\subsubsection*{Emission line profile analysis}
We used the de-reddened and sky-subtracted data cube to extract spectra of the respective ionisation states of Fe. Since the emission lines from the reverse-shocked ejecta are very weak compared to the Balmer dominated H$\alpha$ emission, we used spatial binning to obtain a clearly identifiable signal for curve fitting. We selected several smaller regions from both the eastern and western limb-brightened edges within the white and green rectangles, respectively, in Figure~\ref{fig:rgb} to avoid contamination from stellar emission. For each emission line, we selected and binned individual spaxels with strong S/N from both the eastern and western sides, to enable robust Gaussian profile fitting.
%From Figure~\ref{fig:east} and Figure~\ref{fig:west}, it is evident that [Fe \textsc{xiv}] has the maximum S/N ratio, followed by [Fe \textsc{x}], [Fe \textsc{ix}], and [Fe \textsc{xi}]. In contrast, [Fe \textsc{xv}] has a high S/N ratio on the western side, but it decreases on the eastern side.
We fit the emission lines using a Gaussian profile with a linear continuum, using the \texttt{curve\_fit} function from \texttt{scipy.optimize} (see Figure~\ref{fig:east} and Figure~\ref{fig:west}). The \texttt{curve\_fit} function provided the optimal parameters after fitting, along with the associated covariance, which we used to calculate the errors. We calculated the velocity width of the lines using the formula $V_{\mathrm{width}} = 2.3548\, c\, \sigma_\lambda/\lambda_0$  for the full width at half maximum (FWHM), where $\lambda_0$ is the central wavelength of the signal and $\sigma_\lambda$ is the standard deviation of the Gaussian in wavelength units. Using the outputs from the \texttt{curve\_fit} function for $\lambda_0$ and $\sigma_\lambda$ for all ionisation states of Fe, we computed their respective velocity widths. Uncertainties were calculated by standard propagation of error from the covariance matrix returned by the \texttt{curve\_fit} function.

\subsubsection*{Surface Brightness map of reverse-shocked ejecta}

\begin{figure*}
    \centering
    \includegraphics[width=18cm, trim={80 0 90 0}, clip]{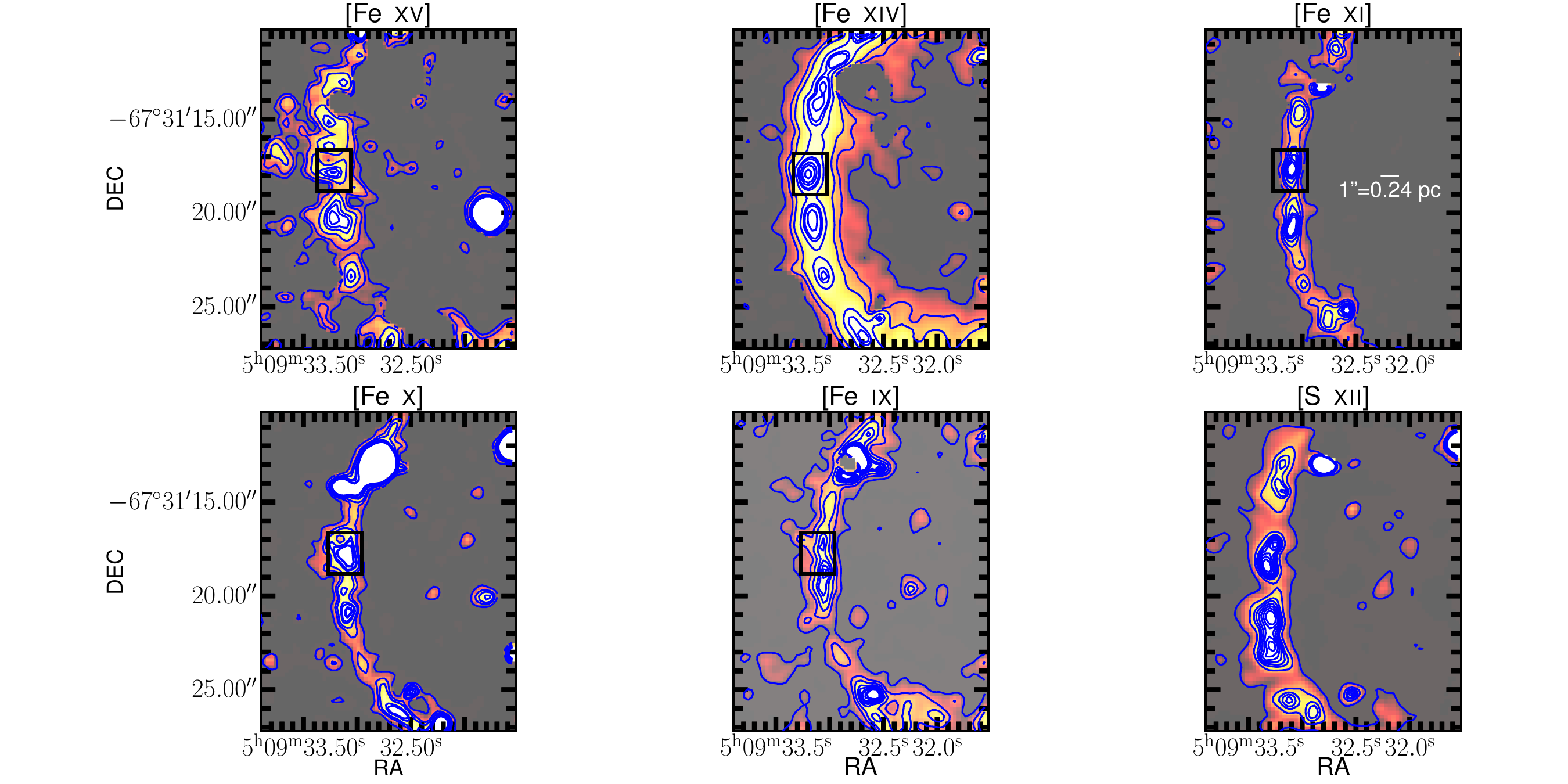}
    \caption{Reverse-shocked ejecta shells of various species are visualised by integrating over their detected emission lines (see Data Analysis). Contour lines in blue with various surface brightness levels are overlaid on the integrated image (see Table.~\ref{tab:table1b} for SB levels). Small circular contours highlight high density clumps/blobs observed in the reverse-shocked ejecta of iron and sulphur. The clump inside the black rectangle is used to visualise the morphological change in Figure~\ref{fig:blob}. }
    \label{fig:iron_clumps}
\end{figure*}

We map the surface brightness (and iso-contours) of the emission lines from the individual ionisation states of Fe and S to interpret the distribution of ejecta material (see Figure~\ref{fig:iron_clumps}). For this, we adopted the following strategies. We smoothed the data cube using Gaussian smoothing with \texttt{kernel} = 2 in \texttt{QFitsView} to improve the visualisation of fainter signals. We selected a region of 68$\times$85 pixels ($13.6''\times 17''$) from the eastern limb-brightened edge of the remnant (white region in Figure~\ref{fig:rgb}), as it exhibits the most prominent clumps with minimal projection effects. We visualised the different species by integrating over the spectral range (between $\lambda_1$ and $\lambda_2$) where each  signal was detected. We integrated two adjacent spectral ranges on both sides of the signal ($\Delta \lambda_{*1}$ and $\Delta \lambda_{*2}$) and subtracted them from the signal to remove any continuum. The selected wavelength ranges are listed in Table~\ref{tab:table1a}. We adopted this process from \texttt{QFitsView}, which allowed us to image the distribution of the ejecta corresponding to their broad emission lines. We present emission maps overlaid with contours at multiple surface-brightness levels (see Table~\ref{tab:table1b}) to enhance the visualisation of high-density clumps, which we identified by closely spaced contours. 
%We displayed the ejecta on a logarithmic scale with an opacity of $\alpha = 0.6$, and overplotted the corresponding contour lines using \texttt{Matplotlib}.

\begin{figure}
    \centering
    \includegraphics[width=11cm, trim={0 0 0 0}, clip]{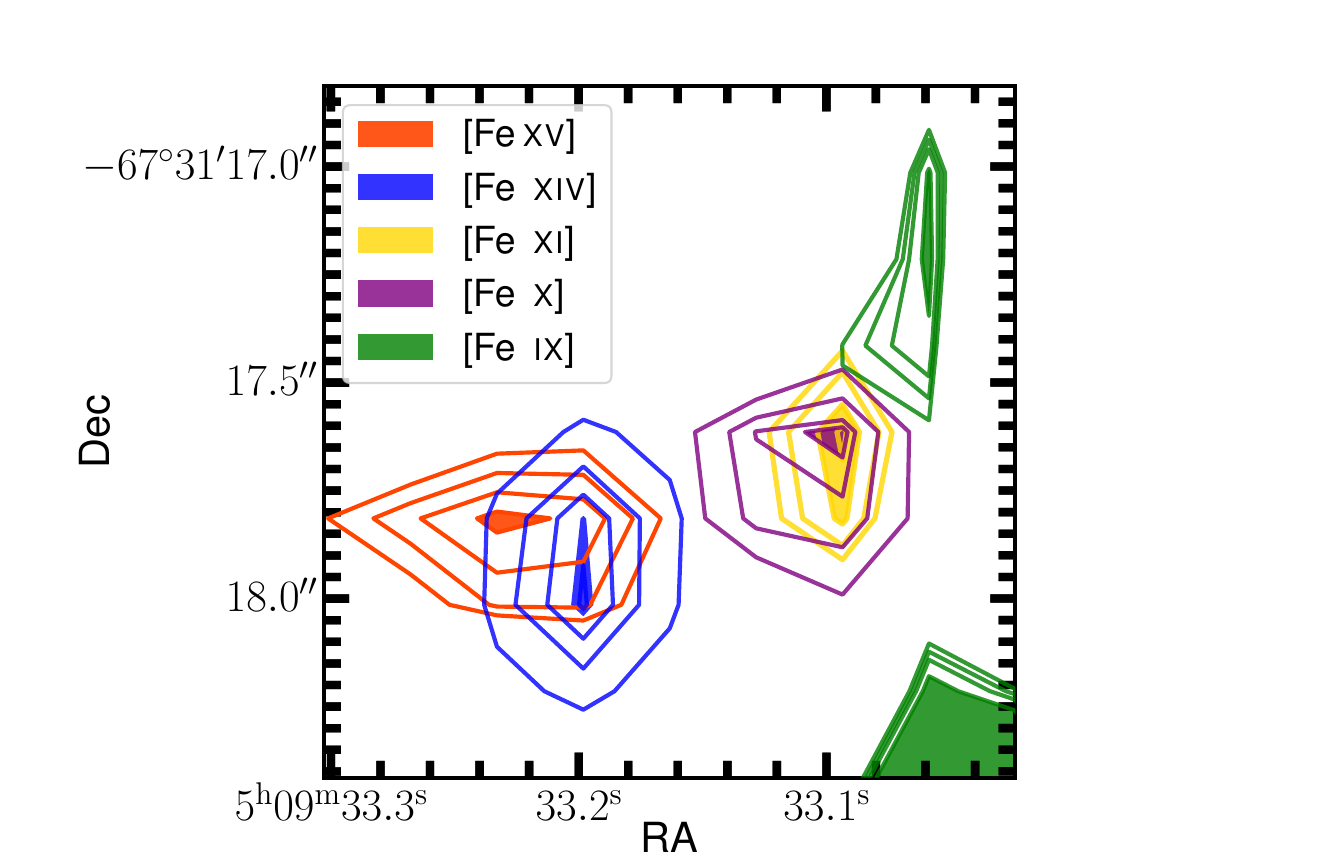}
    \caption{The clump highlighted in Figure~\ref{fig:iron_clumps} is isolated and shown with contour levels across the successive ionisation zones of Fe to track the morphological changes.}
    \label{fig:blob}
\end{figure}

The perfectly circular, saturated (in white) contour lines in Figure~\ref{fig:iron_clumps} in [Fe \textsc{x}], [Fe \textsc{xv}], and [S \textsc{xii}] result from the presence of a star with signal at the same wavelength as the broad emission lines. We selected the ejecta clump highlighted in Figure~\ref{fig:iron_clumps}, between $-67^{\circ}31'19''$ and $-67^{\circ}31'23''$, to study the dynamics and evolution of the clump through successive ionised states of Fe. We present the same clump across all ionised states of Fe together in a single frame, as shown in Figure~\ref{fig:blob}. We coloured the highest contour level in each ionisation state to better track the evolution of the clump and its behaviour under the effect of the reverse shock.
%Fig~\ref{fig:iron_clumps}, the closed near-circular contours represents the clumpy ejecta surrounded by smoother ejecta shell represented by long vertical contour lines. The surface brightness of the plotted contour lines for each emission line is also listed in Table 2. 

%Since the reverse shock ionizes the ejecta in successive zones, the [Fe \textsc{xv}] shell is expected to lie closer to the contact discontinuity (or ejecta-envelope boundary), while the [Fe \textsc{ix}] shell, being the least ionized, should lie further inward. As mentioned in IS19, observation of the optical coronal lines reveals the Doppler shift, spectral broadening, and also the ejecta structure.
\subsubsection*{Radial profiles of coronal Fe emission}
To resolve the radial stratification of coronal Fe ionisation stages across the limb-brightened eastern edge of SNR 0509--67.5, we extracted one-dimensional surface brightness profiles from the de-reddened, Gaussian-smoothed MUSE datacube. For each line we first constructed a continuum-subtracted emission map as described above. For each map, we extracted a 3$\times$27-spaxel rectangular aperture on the limb-brightened eastern edge (white rectangle in Figure~\ref{fig:rgb}), with its long axis oriented radially outwards along right ascension. Summing over  declination gives the radial profile and its uncertainty. We then convert the spectral flux to surface brightness using the MUSE flux scaling of $10^{-20}~\mathrm{erg~s^{-1}\,cm^{-2}}$\,\AA$^{-1}$ (see Figure~\ref{fig:line_profile}).

%We applied a \texttt{Python} routine to compute the line profiles of these ionized ejecta which does the following: we selected the limb-brightened edge of the remnant for this purpose, using a region of 68$\times$85 spaxels highlighted by a white rectangle in Figure~\ref{fig:rgb}. We selected each ionisation state of Fe using a method similar to that described below. We computed the line profiles for the respective ionized Fe species by integrating the flux over rectangular apertures spanning along the y-axis ($3.2~\mathrm{arcsec}$) and plotting along the x-axis ($5.6~\mathrm{arcsec}$) within the highlighted region on the eastern side of the remnant. We plotted the surface brightness (SB) of each coronal emission line and annotated the position of the peak in the graph, which marks the position of each ionized species of Fe. The separation between [Fe \textsc{xv}] and [Fe \textsc{xiv}] is 0.05 pc, between [Fe \textsc{xiv}] and [Fe \textsc{xi}] is 0.1 pc, between [Fe \textsc{xi}] and [Fe \textsc{x}] is 0.09 pc, and between [Fe \textsc{x}] and [Fe \textsc{ix}] is 0.05 pc.
\section{Results}
IS19 reported the first observation of optical coronal lines from the reverse-shocked ejecta of young supernova remnants. They used the ejecta shells as a diagnostic tool to constrain the explosion energy and mass of the progenitor. IS19 previously detected and reported [Fe \textsc{xiv}] 5302.86 \AA\ and [Fe \textsc{xv}] 7059.59 \AA\ emission lines in SNR 0509 and N103B, using a MUSE data cube with a total exposure of 5608 s. Our new deep MUSE cube, with an on-source integration time of ${\sim}105300$ s (29 h 15 min), not only confirms the presence of [Fe \textsc{xiv}] and [Fe \textsc{xv}] but also reveals additional ionisation stages of Ca, S, and Fe in the ejecta, including [Fe \textsc{xi}] 7894.06 \AA, [Fe \textsc{x}] 6374.54 \AA, and [Fe \textsc{ix}] 8236.55 \AA\ \citep[all wavelengths measured in air, as given by][]{delzanna2018}.
\subsection*{Ionisation in the reverse-shocked ejecta}
\begin{figure*}
    \includegraphics[width=16cm, trim={90 20 130 70}, clip]{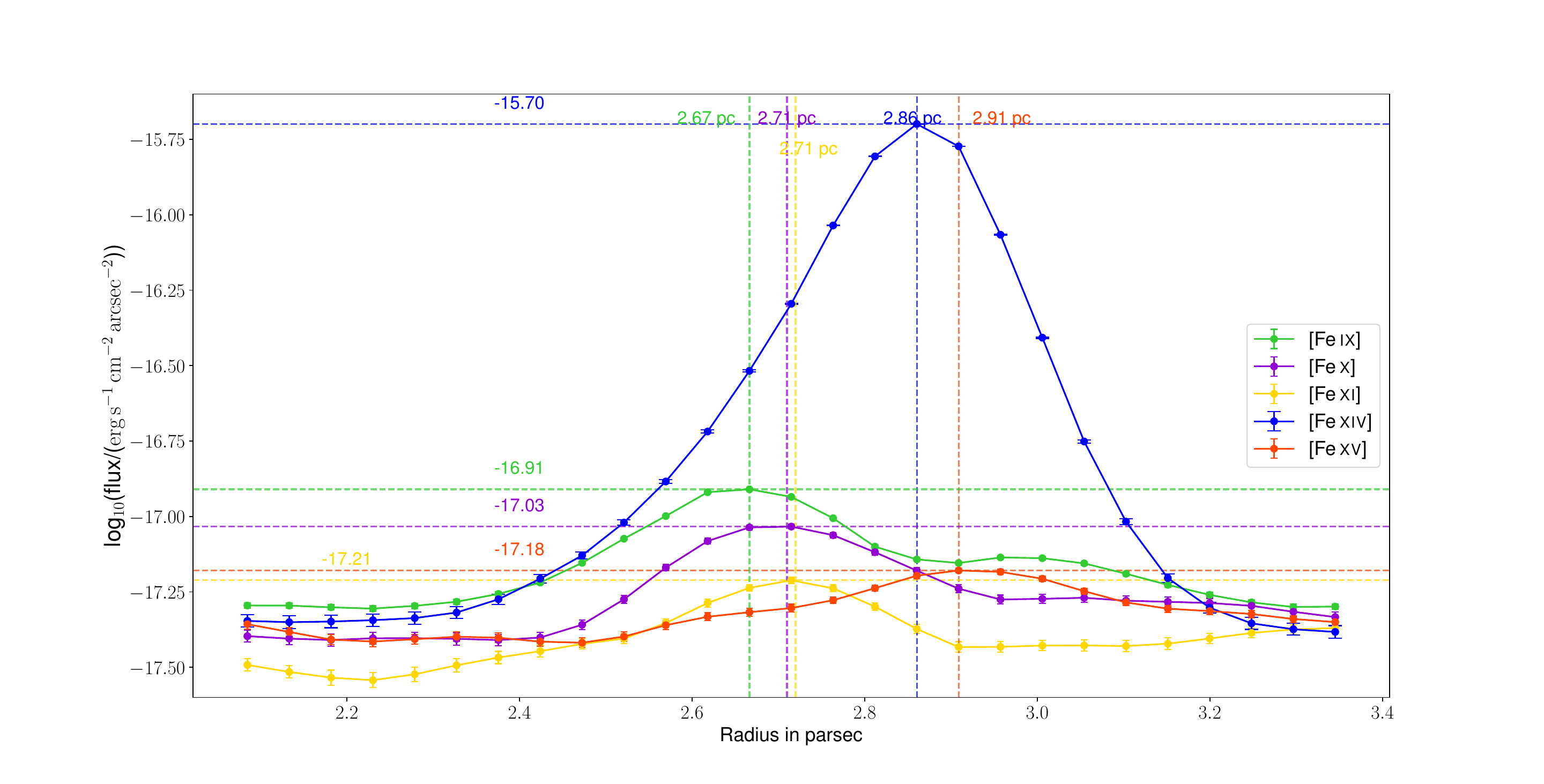}
    \caption{Observed surface brightness (in log scale) of forbidden coronal emission lines from various ionised states of Fe in the reverse-shocked ejecta of SNR 0509. Also indicated are  the radii at which each detected ionisation state reaches its peak, namely [Fe \textsc{xv}], [Fe \textsc{xiv}], [Fe \textsc{xi}], [Fe \textsc{x}], and [Fe \textsc{ix}]. The spatial offsets between the limb-brightened shells of these ionised Fe species in the reverse-shocked ejecta are marked above each profile, and their corresponding maximum surface brightnesses are shown on the left. }
    \label{fig:line_profile}
\end{figure*}

\begin{figure*}
    \includegraphics[width=9cm, trim={60 40 40 60}, clip]{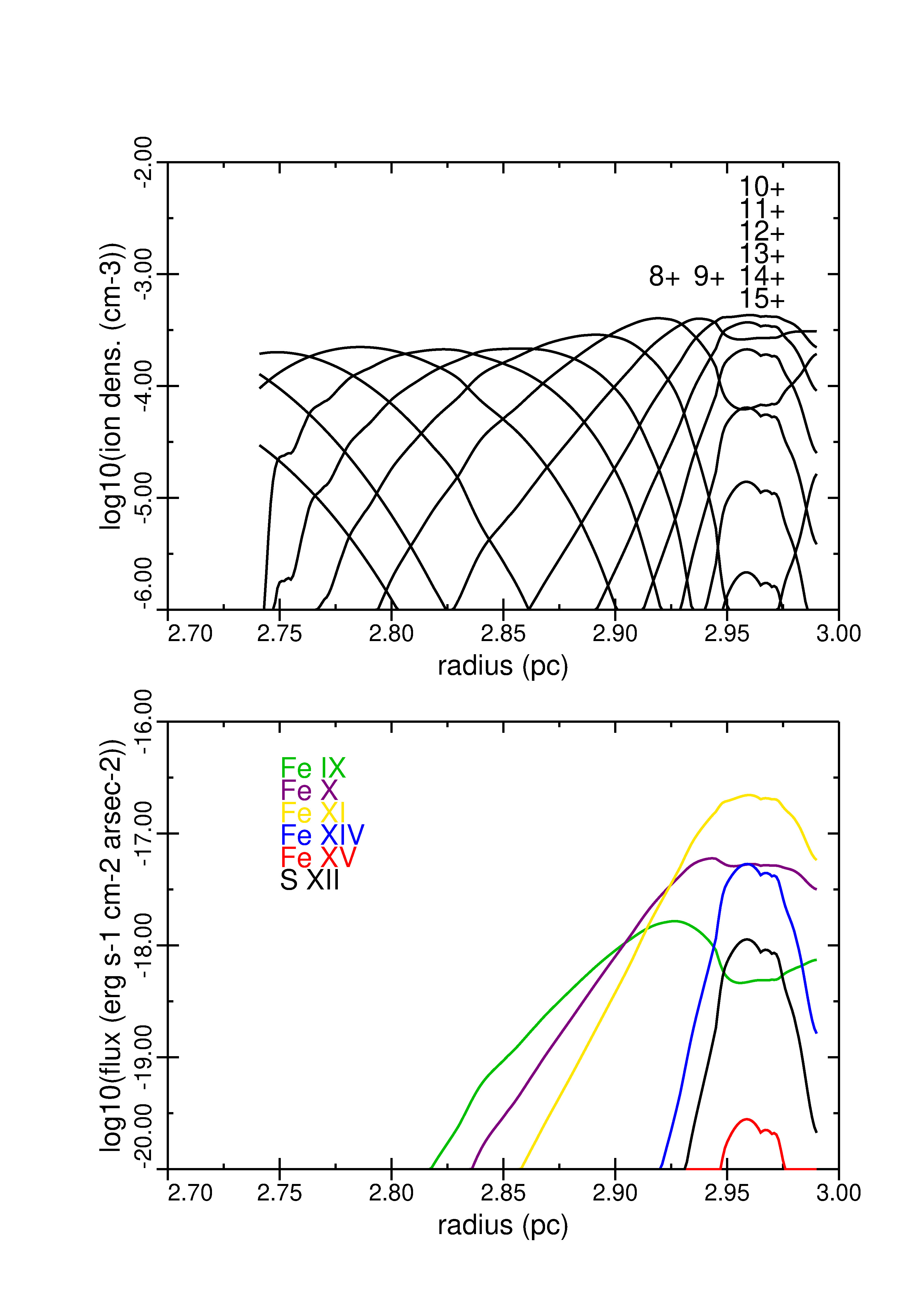}
    \includegraphics[width=9cm, trim={60 40 40 60}, clip]{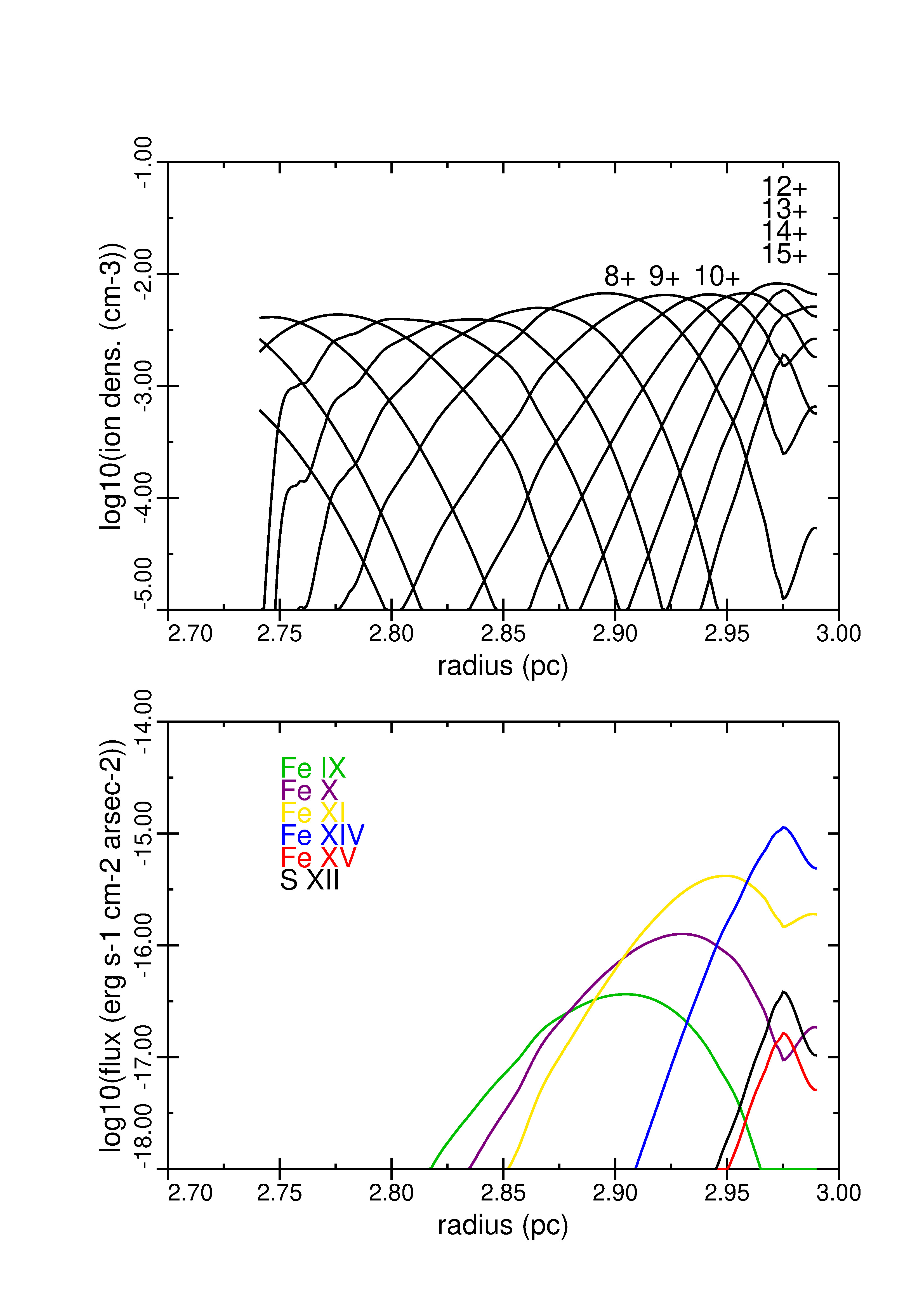}
    \caption{1D analytical model to predict the occurence of highly-ionised coronal lines in the reverse-shocked ejecta of a SNR originating from a sub-Chandrasekhar mass white dwarf explosion. Top: Number density of each charge state fraction of Fe and their predicted radii from the remnant's centre where they emerge for model A (left) and model B (right). Bottom: Predicted surface brightness (in log scale) of the coronal emission lines for each Fe ionisation states for model A (left) and model B (right). }
    \label{fig:nei_lines}
\end{figure*}

We create 1D SNR models \citep[using][]{Truelove1999} where the reverse shock position is computed analytically using the specific explosion parameters summarised for SNR 0509 in IS19 -- ejecta mass $M_{ej}=1.0~\mathrm{M}_\odot$, explosion energy $E=1.5\times10^{51}$ erg, and age $t=310$ years assuming homologous expansion. The model further computes the non-equilibrium ionisation of Fe in the ejecta following the reverse shock passage as in \citet{Laming2003} and IS19. For the model above we assume two cases, A and B, with an ejecta density given by an outer envelope power law $v^{-7}$ for ejecta expansion velocity $v$ and a uniform density ejecta core. The composition of the ejecta is set by O:Si:S:Ca:Fe = 0.2:0.5:0.2:0.05:0.05 for case A and O:Si:S:Ca:Fe = 0.2:0.5:0.2:0.05:1.05 for case B. Case B has a local overdensity by a factor of two, coming from an increased mass of Fe in attempt to reproduce the strength of the Fe lines observed in SNR 0509.
The \citet{Truelove1999} model remains the same, except that when encountering the increased density in the ejecta clump the reverse shock velocity is reduced in accordance with the model in \citet{Sgro1975}. 
The model computes the charge state fractions of Fe for both cases at different radii (top panel of Figure~\ref{fig:nei_lines}). The radial extent of these ionisation states of Fe roughly matches with the radial extent of the same ionisation states of Fe observed in SNR 0509. The peak radius of each ionisation state is taken as the position of maximum surface brightness with the half-spaxel sampling setting a positional uncertainty of $\sim$$0.024$~pc. The separation between [Fe \textsc{xv}] and [Fe \textsc{xiv}] is 0.05 pc, between [Fe \textsc{xiv}] and [Fe \textsc{xi}] it is 0.15 pc,  [Fe \textsc{xi}] and [Fe \textsc{x}] are unresolved, and between [Fe \textsc{x}] and [Fe \textsc{ix}] the separation is 0.04 pc. The bottom panel of Figure~\ref{fig:nei_lines} shows estimates of the surface brightness (SB) of specific coronal forbidden lines originating from various ionisation states of Fe for the two different ejecta compositions. The line emissivities are calculated from atomic data given by \citet{DelZanna2014} for Fe \textsc{ix} (865 levels), \citet{DelZanna2012} for Fe \textsc{x} (552 levels), \citet{DelZanna2013} for Fe \textsc{xi} (996 levels), \citet{Aggarwal2014} for \textsc{xiv} (136 levels), \citet{Fernandez2014} for Fe \textsc{xv} (53 levels), and \citet{Zhang1994} for S \textsc{xii} (15 levels). All models include full radiative cascades. We also include estimates of
the impact excitation rates by heavy ions, typically O$^{6+}$ following \citet{Seitenzahl2019} using the approach of \citet{Laming1996} 
generalised from E1 to M1 transitions,
for the ions with ground state fine structure transitions, Fe \textsc{x}, Fe \textsc{xi}, Fe \textsc{xiv},
and S \textsc{xii}. This can increase forbidden line emissivities over those calculated from the electrons alone by a factor of up to 2. Leveraging the superior spatial resolution of MUSE ($0.2'' \times 0.2''$), we mapped the radial ejecta profiles of five Fe ionisation states -- [Fe\,\textsc{xv}], [Fe\,\textsc{xiv}], [Fe\,\textsc{xi}], [Fe\,\textsc{x}], and [Fe\,\textsc{ix}] -- and found their peak emission regions to lie between 2.6 and 2.91 pc (Figure~\ref{fig:line_profile}).

In Figure~\ref{fig:line_profile}, [Fe\,\textsc{xiv}] is the brightest coronal line, followed by [Fe\,\textsc{ix}], with [Fe\,\textsc{xv}] being the faintest. To reproduce the observed surface brightness (SB) of the ionised Fe species, we adopted the ejecta compositions described above. The analytically computed SB values for the Fe ionisation states derived from case A, shown in the left panel of Figure~\ref{fig:nei_lines}, exhibit substantial discrepancies with all of the observed lines except [Fe\,\textsc{x}]. In particular, the model predicts a much lower SB for [Fe\,\textsc{xiv}] and a higher SB for [Fe\,\textsc{xi}], in clear opposition to the observations. Nonetheless, the predicted SB of [Fe\,\textsc{x}] in case A agrees well with the observed line profiles. 

The SB values of the Fe ionisation states computed from case B, displayed in the right panel of Figure~\ref{fig:nei_lines}, agree well with the observations for [Fe\,\textsc{xiv}], [Fe\,\textsc{xv}] and [Fe\,\textsc{ix}]. However, case B also has limitations: it fails to match the observed SB of [Fe\,\textsc{x}] and [Fe\,\textsc{xi}], for which it predicts values that are higher by several orders of magnitude (See Table~\ref{tab:line_fluxes}). This possibly suggests that we are looking at a ``two ring'' structure in Fe, similar to that seen in [Ca\,\textsc{xv}] \citep{Das2025}. Further discussion of such ideas is beyond the scope of this paper, but much motivation exists for future modeling.

\begin{table}
\centering
\caption{Comparison of observed and model-predicted Fe line surface brightness at the eastern limb brightened edge, integrated over the radial extent of each ionisation shell in Figure~\ref{fig:line_profile}. Observed uncertainties are the 1$\sigma$ propagated errors. Model A and Model B correspond to the two ejecta compositions described in Section~3.}
\label{tab:line_fluxes}
\begin{tabular}{lccc}
\hline
Ion & Observed & Model A & Model B \\{}
    & \multicolumn{3}{c}{(erg s$^{-1}$ cm$^{-2}$ arcsec$^{-2}$ pc)} \\{}\\
\hline
{[Fe\,\textsc{ix}]}  & $(8.97\pm0.02)\times10^{-18}$ & $1.09\times10^{-19}$ & $2.59\times10^{-18}$ \\
{[Fe\,\textsc{x}]}   & $(7.04\pm0.04)\times10^{-18}$ & $3.82\times10^{-19}$ & $8.34\times10^{-18}$ \\
{[Fe\,\textsc{xi}]}  & $(5.07\pm0.03)\times10^{-18}$ & $9.74\times10^{-19}$ & $2.75\times10^{-17}$ \\
{[Fe\,\textsc{xiv}]} & $(4.64\pm0.05)\times10^{-17}$ & $1.35\times10^{-19}$ & $5.55\times10^{-17}$ \\
{[Fe\,\textsc{xv}]}  & $(6.08\pm0.03)\times10^{-18}$ & $6.38\times10^{-22}$ & $7.29\times10^{-19}$ \\\\
\hline
\end{tabular}
\end{table}

\subsection*{Variation of reverse shock velocity}
The reverse-shocked ejecta produces broad optical coronal emission lines, and the velocity width of these lines gives us an indirect measurement of the reverse shock speed. The interaction of the reverse shock with the ejecta also heats up the ejecta, contributing to thermal motions of the ions, which in turn contributes to thermal broadening \citep{Vink2012}. Broadening of the emission lines can also occur due to the introduction of turbulence in the ejecta as a result of the ejecta-reverse shock interaction \citep{McKee1995}. Thus, it is fair to state that the broadening of the coronal emission lines depends directly on the effect of the reverse shock, and the velocity width of the emission lines can be associated with the reverse shock velocity, as both the ionisation and thermal effects are directly dependent on the velocity of the shock.\\

\begin{figure*}
    \centering
    \includegraphics[width=7cm, trim={0 0 45 20}, clip]{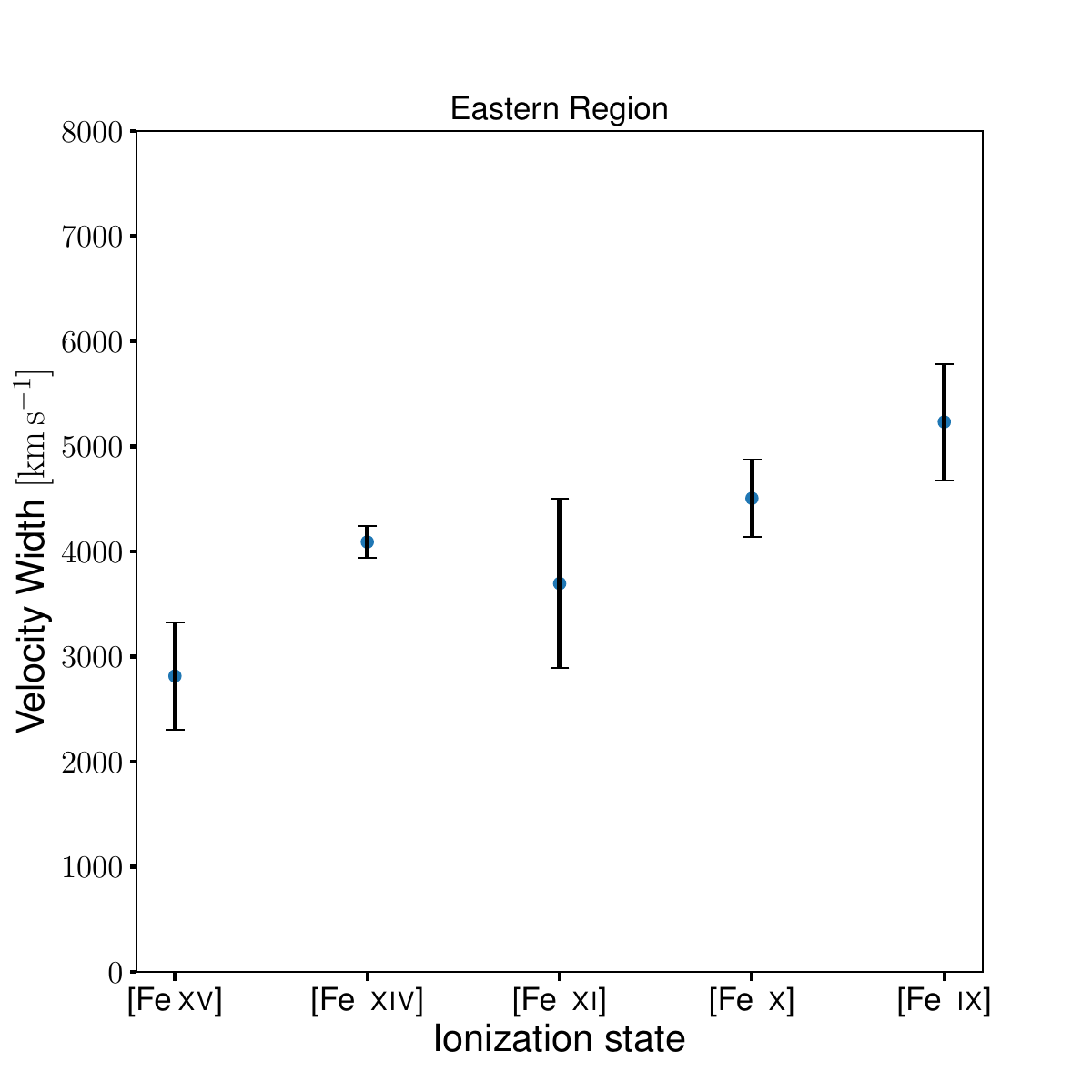}
    \includegraphics[width=7cm, trim={0 0 45 20}, clip]{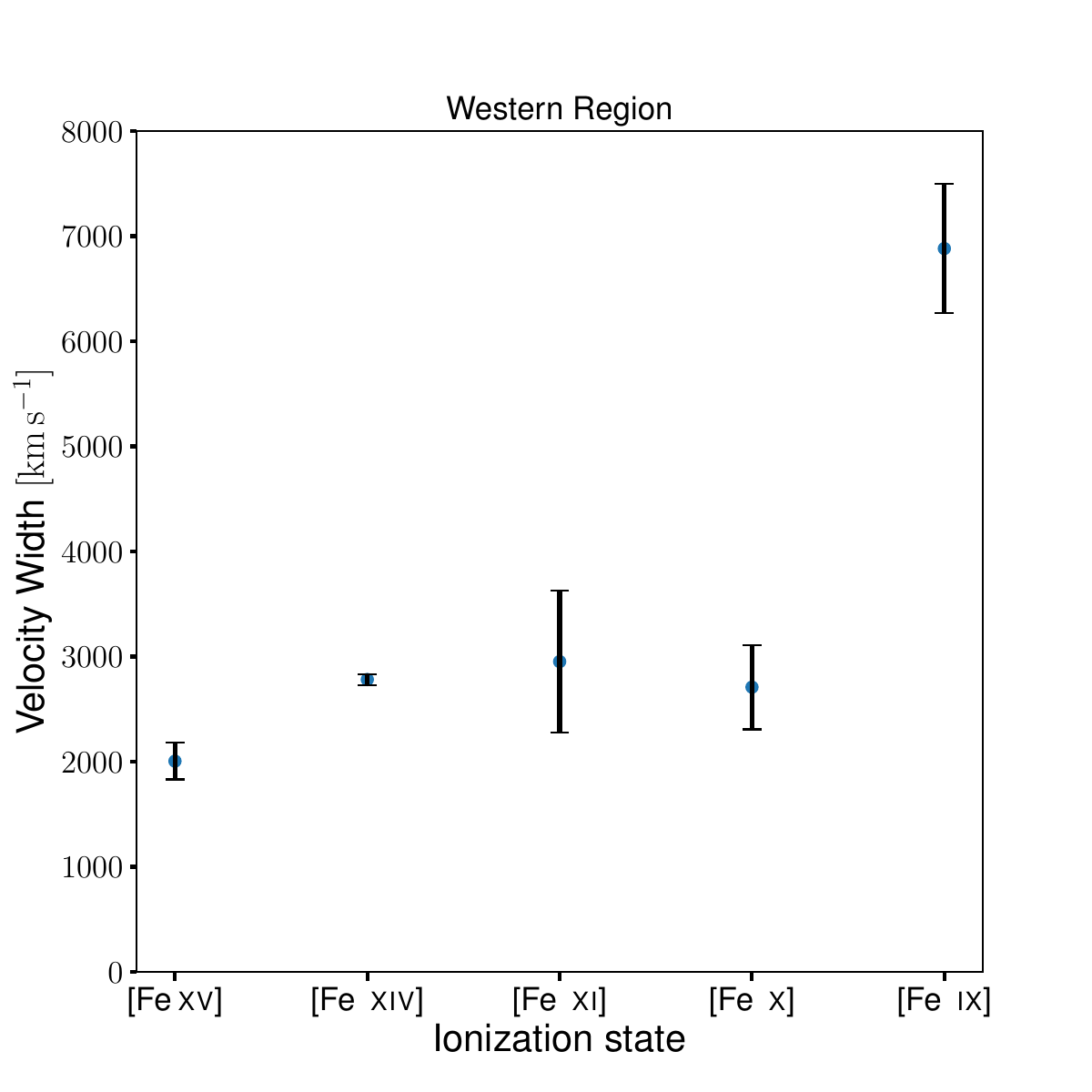}
    \caption{Velocity widths of the ionised Fe states with standard propagated uncertainties (1 $\sigma$)} shown as error bars. Left: Velocity widths derived from the Gaussian fits shown in Figure~\ref{fig:east}. Right: Velocity widths derived from the Gaussian fits shown in Figure~\ref{fig:west}. In both regions, [Fe\,\textsc{xv}] exhibits the narrowest widths, while [Fe\,\textsc{ix}] shows the broadest.
    \label{fig:east_trend}
\end{figure*}

As SNR 0509 is a young Type Ia remnant, transitioning from free-expansion to the Sedov-Taylor phase, the reverse shock velocity is expected to increase as it traverses inwards \citep{chevalier1982self,Leahy2017}. As described earlier, we fit single component Gaussian profiles to determine the velocity widths of the coronal emission lines of the different ionisation states of Fe. Figure~\ref{fig:east_trend} illustrates that, in both the eastern and western regions, the velocity width progressively increases as the ionisation state of Fe decreases. The western region exhibits a more pronounced rise in velocity widths compared to the eastern region, likely because it is interacting with a denser CSM. Figure~\ref{fig:rgb} presents a composite image combining H$\mathrm{\alpha}$ (orange), [Fe \textsc{xiv}] (magenta), and [S \textsc{xii}] (cyan), revealing pronounced interaction and structural irregularities on the western side of the remnant. This behavior is most likely caused by the expanding ejecta encountering a dense surrounding medium. The denser ambient medium on the western side impedes forward shock expansion, generating a stronger reverse shock that propagates inward more rapidly in that region. 

%For $\gamma = 5/3$ gas, the line full width at half maximum (FWHM) is very close (within a few percent) of the reverse shock velocity, assuming negligible equilibration. The model predictions for Fe \textsc{ix} of 5011-3644 km s$^{-1}$ and for Fe \textsc{xiv} and Fe \textsc{xv} of 4805-3500 km s$^{-1}$, where the quoted ranges account for unperturbed density to two times overdense, compared with Figure~\ref{fig:east_trend} indicate very good agreement for the eastern region, for which the model is designed, with a suggestion of stronger clumping in the outer Fe ejecta, [Fe \textsc{xv}], as should probably be expected.

\subsection*{Clump formation in the SNR 0509}

We report the detection of distinct clumpy regions in all observed Fe ionisation states and [S\,\textsc{xii}] within the eastern limb-brightened ejecta. As the reverse shock propagates inwards, the outermost layers of the ejecta interact first and exhibit higher ionisation compared to the freshly ionised ejecta closer to the center of the remnant. Thus, [Fe\,\textsc{xv}] appears in the outermost layers, while [Fe\,\textsc{ix}] originates from deeper layers. Figure~\ref{fig:iron_clumps} shows contour maps of different levels of observed optical flux (see Table~1) for the various successively ionised Fe species in the ejecta, along with an overlay of their distribution map. Successive Fe ionisation states are shown together for direct comparison and highlight the radial extent of the shells along with their distinct morphology. Some clumps are more prominent and their existence can be traced from [Fe\,\textsc{ix}] to [Fe\,\textsc{xv}] (e.g., the clump highlighted in the black square).
%A detailed comparison across ionisation states shows that five Fe-rich clumps are consistently visible from [Fe \textsc{ix}] through [Fe \textsc{xv}], suggesting persistent spatial structure. A closer analysis reveals that the clumps are surrounded by a smoother ejecta. The circular contour with a white center at the top of both [Fe \textsc{x}] and [Fe \textsc{ix}] is due to emission from bright background stars at the same wavelength.

The clumps observed in [Fe \textsc{xv}] and [Fe \textsc{ix}] likely trace the same ejecta structures at different ionisation stages following the passage of the reverse shock. To examine their morphological evolution, we isolated one representative ejecta clump (black box in Figure~\ref{fig:iron_clumps}); Figure~\ref{fig:blob} shows its evolution from [Fe \textsc{ix}] to [Fe \textsc{xv}]. The observed morphological progression of individual clumps (compact in low-ionisation maps; broader in high-ionisation maps) can be explained via a combination of physical processes that act as the reverse shock penetrates the ejecta. First, direct shock compression increases local temperature and ionisation, most strongly at a knot; simultaneously, shear at the interface of the ejecta boundary drives Kelvin–Helmholtz growth, while RM/RT modes fragment the knot as it is decelerated \citep{Wang2001,Blondin2001a, Wang2011a}. These instability-driven processes naturally produce the appearance of a compact core with surrounding filamentary or fragmented structure in more early shocked ejecta. Second, internal pressure variations produced early by radioactive heating (the so-called Ni-bubble effect) can induce alternating compression and rarefaction of ejecta clumps. Third, microphysical effects such as thermal conduction, radiative cooling, and local magnetic fields modify the clump’s survivability and morphology: magnetic tension can suppress small-scale fragmentation and produce elongated morphologies, while efficient conduction and ablation tend to smear out small-scale structure \citep{Cowie1977,Blondin2001b, Vieser2007, Orlando2008}. %Finally, projection and ionisation-selection effects (each coronal line samples a different temperature/ionisation window) cause the same three-dimensional structure to appear differently in different lines.

Taken together, the simplest consistent picture is that we are observing the growth of Rayleigh–Taylor (RT) instabilities, in which dense ejecta knots are progressively processed by an inward-moving reverse shock. The knot cores largely survive (hence detection in multiple ions), while the outer layers are stripped and fragmented, producing broader, more structured emission in the highest-ionisation lines. We emphasise that while this interpretation is strongly suggested by the morphology and line width trends, disentangling the relative importance of Ni-bubble compression versus instability-driven fragmentation will require either targeted hydrodynamic modeling or further multi-wavelength constraints (e.g., X-ray counterparts, polarisation, or resolved PV diagrams). Until such tests are available, statements about Ni-bubble dominance and other microphysical effects are framed as a plausible mechanism rather than a firm conclusion.

%The successive ionized states of Fe and the deep MUSE data again reveal the effect of reverse shock interaction in shaping the morphology of high-density clumps in the ejecta formed due to hydrodynamical instabilities. It is also hypothesized that possible nickel bubble expansion can cause alternating compression and rarefaction of ejecta clumps, altering their morphology over time \citep{Wang2005, Wang2008}. The ejecta clump in [Fe \textsc{ix}] shown in Figure~\ref{fig:blob} appears compressed and vertically oriented within the shell. With the expansion of the shell, the ejecta clumps begin to expand.

[S\,\textsc{xii}] also appears to exhibit clumps in the ejecta of SNR~0509. However, the clumps exhibited by [S\,\textsc{xii}] differ from those seen in Fe. The [S\,\textsc{xii}] emission reveals fewer but significantly more distinct and sharper clumps compared to those observed in the ionised Fe maps. However, they are also surrounded by comparatively smoother ejecta, similar to Fe. Across all observed species, the clumps consistently appear embedded in limb-brightened shell structures, regardless of ionisation state or element. This represents the first high-resolution optical observation of clumps in the ejecta of a Type Ia SNR. Ionisation ages fitted to X-ray spectra also indicate clumping in 0509 \citep{Warren2004} and in SN 1006 \citep{uchida2013,Laming2026}, generally to a higher degree in the non-Fe ejecta than suggested here.

\begin{figure*}
    \centering
    \includegraphics[width=15cm, trim={40 0 40 0}, clip]{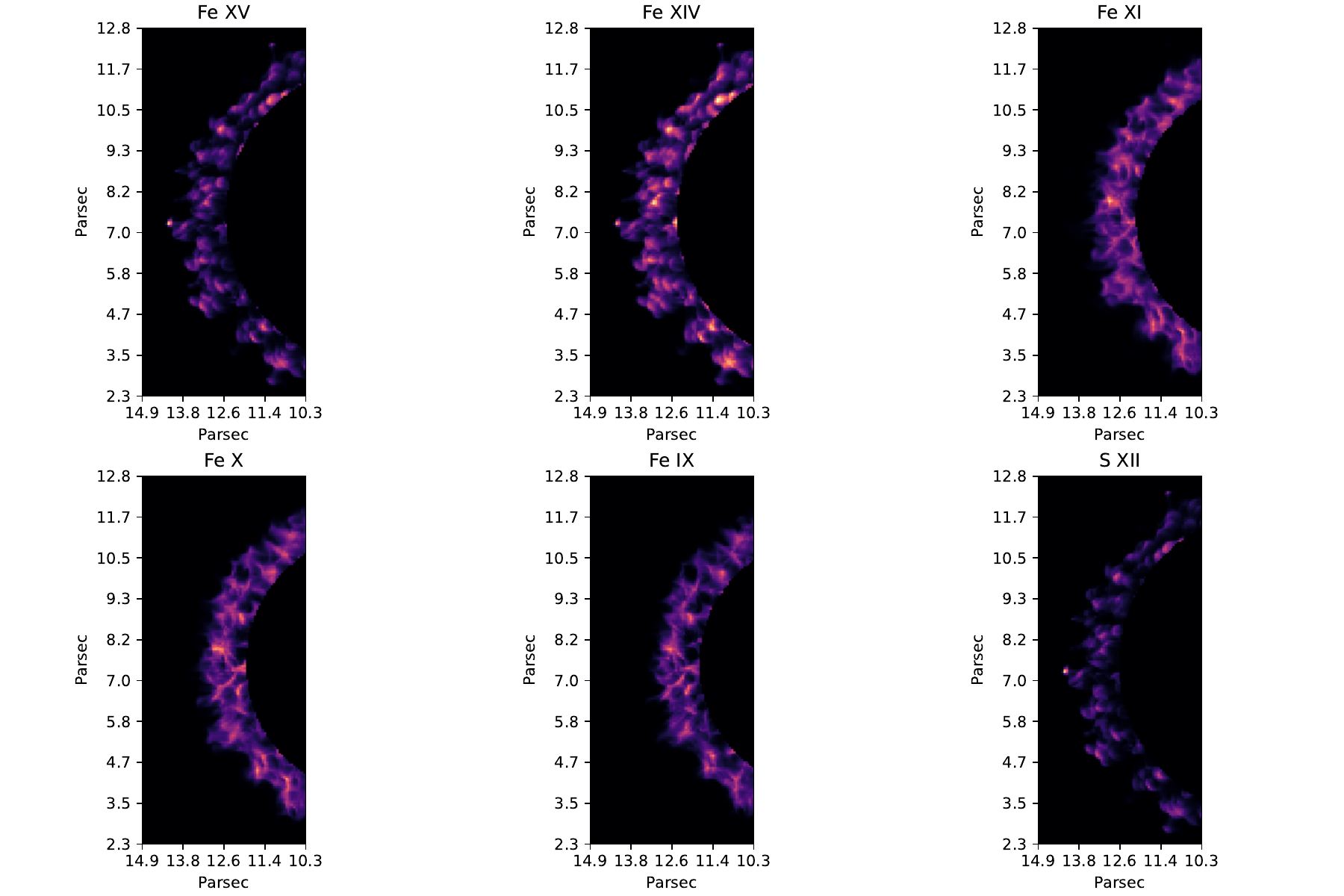}
    \caption{Slices from the D6 model showing the ejecta distribution of highly ionised Fe and S species, along with clumping, similar to the ionisation stages observed in SNR 0509. The inner region of the remnant is masked by excluding all ejecta within a radius set to $\sim$1.5 times the inner radius measured for the individual ionised shells in the observations, thereby reproducing the observed limb-brightened morphology.}
    \label{fig:modelclump}
\end{figure*}

\begin{figure*}
    \centering
    \includegraphics[width=15cm, trim={20 20 20 20}, clip]{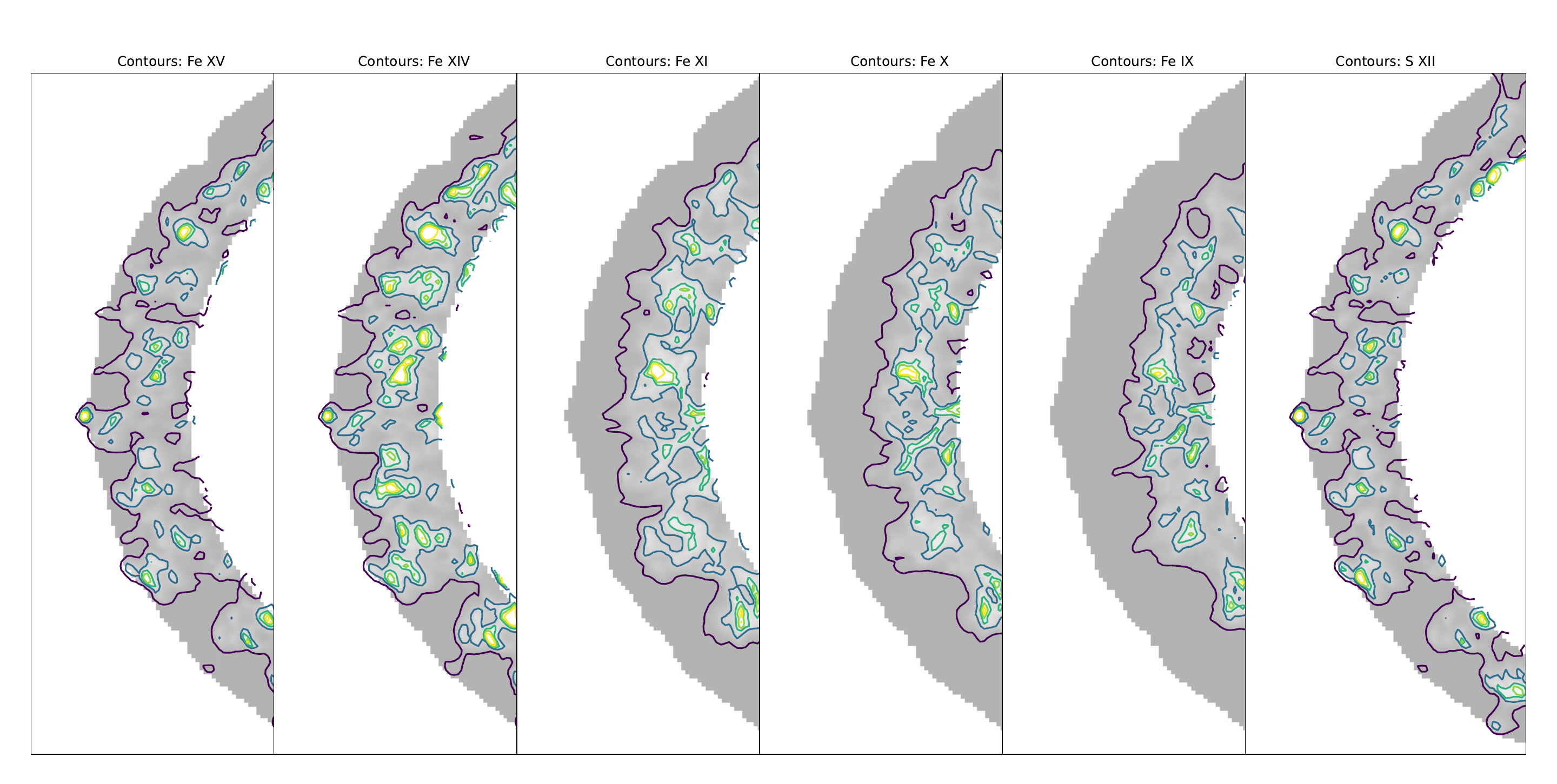}
    \caption{The ejecta of the D6 model with various ionised states of Fe and [S\,\textsc{xii}] are shown with contour lines to distinguish the high density clumps/blobs from the smoother ejecta in the model.}
    \label{fig:modelcontour}
\end{figure*}

%The initiation of reverse shock is self-explanatory of the fact that the supernova ejecta expansion is slowing down as the forward shock sweeps enough surrounding material. This in turn marks the time frame for the development of hydrodynamic instabilities, as the high density ejecta of SNR 0509 interacts with a lower density ambient gas. The RM instability causes density features in the initial SNR system, such as modulations in the radius of the inner and outer edges of the CSM and ejecta, to grow. 
We employed the 3D hydrodynamical remnant-forward modelled version of the dynamically driven double degenerate double detonation (D6) explosion mechanism (see \citet{Tanikawa2018} for details of the explosion mechanism and \citet{Ferrand2022} for the remnant model), motivated by recent results of SNR 0509 originating from a double WD merger system \citep{Das2026b}. The dynamical age of the D6 remnant at 500 years matches that of SNR 0509, implying that both are at similar stages of evolution (SNR 0509 is expanding 1.3-1.7 times slower than the model). Figure~\ref{fig:modelclump} shows similar plots of the morphology of the ejecta at the eastern limb-brightened edge of the D6 model. We show the ejecta distribution of similar ionisation stages of Fe and S for direct comparison with our observations. Figure~\ref{fig:modelcontour} shows the iso-contour levels in the D6 model to highlight localised high-density blobs, analogous to the observed clumps in SNR 0509. The radial extent of the ionised ejecta shells is not directly comparable to our observations, as the model evolves faster than SNR 0509 due to differences in the ambient density. As a result, the full spatial extent of the model remnant is larger by a factor of $\sim$1.3–1.7 in expansion rate (corresponding to up to a factor of $\sim$3.4 in diameter), even though the dynamical ages are similar. [Fe\,\textsc{xiv}] from the D6 model shows distinct localised high-density clumps in Figure~\ref{fig:modelcontour}, similar to those observed in Figure~\ref{fig:iron_clumps}. [Fe\,\textsc{xv}] and [S\,\textsc{xii}] trace the ejecta that were shocked earliest and have therefore resided longest at the contact discontinuity, where Rayleigh–Taylor fingers continue to grow. The greater fragmentation seen in these species in the model is thus more plausibly a consequence of longer RT growth time. This effect can be seen in the [Fe\,\textsc{xv}] ejecta in the model and, to some extent, in the observations. However, [S\,\textsc{xii}] in the ejecta of SNR 0509 deviates from the model, with smoother ejecta and more distinct clumps observed.


\section{Conclusions}
In this paper, we present a comprehensive deep optical observational study of the young Type Ia SNR 0509 in the Large Magellanic Cloud using MUSE on the VLT, along with a comparison of the observations with available 1D and 3D models. We confirm that Type Ia SNRs can reveal non-radiative reverse-shocked ejecta in the optical through collisional ionisation and excitation, as first established by \citet{Seitenzahl2019}. Building on this discovery, our deep MUSE observations provide a spatially resolved view of these coronal emission lines across the remnant, allowing us to map zones and clumps of successive ionisation states in iron, from [Fe\,\textsc{xv}] to [Fe\,\textsc{ix}] and [S\,\textsc{xii}], which appear as limb-brightened clumpy shells. This enables morphological comparison with 3D hydrodynamical remnant models, at least qualitatively for now. The high spatial resolution of our data reveals the clumpy substructure within these shells and allows us to trace the morphological evolution of ejecta clumps across different ionisation states.

%The reverse-shocked ejecta create zones of successive ionisation states due to non-equilibrium ionisation driven by reverse shock interaction. Collisional excitation of these charge states produces faint broad coronal lines that directly trace the ejecta. Since these emission lines are faint, a $\sim$30 h observation is required to reveal the limb-brightened edges of these ionized species, which appear as ejecta shells at specific distances from the centre. Our observations of these limb-brightened ejecta shells, along with mapping their line profiles, enable us to determine the precise location and surface brightness distribution of their extent from the centre of the remnant.

We compare our 1D analytical models (Model~A and Model~B) to the observed surface brightness of the Fe coronal lines in SNR 0509 (see Figure~\ref{fig:nei_lines}). Model A, with no overdensity, fails to reproduce most of the observed coronal lines except [Fe\,\textsc{xiv}]. Model B, with its Fe mass fraction increased from 0.05 to 0.525 and concurrent factor of two increase in total mass density, successfully reproduces the high surface brightness of [Fe\,\textsc{xiv}] and [Fe\,\textsc{ix}] as well as the lower surface brightness of [Fe\,\textsc{xv}]. In spite of the good agreement, we acknowledge that Model B is not fully satisfactory due to several caveats: the unknown cross-ionisation effects between different ionised species and the possibility that the ejecta is not uniformly denser by a factor of two. Nevertheless, the model comparison indicates the need for enhanced density or clumping in the Fe ejecta to reproduce the significantly higher surface brightness of [Fe\,\textsc{xiv}] and a possible ``two-ring" structure for Fe with higher density. The calculated radii at which the surface brightnesses of the ionised Fe shells peak are broadly consistent with the observations. However, the exact shell radii in SNR 0509 are influenced by the density of the ambient medium, its interaction with the ejecta, and the development of Rayleigh–Taylor instabilities. As this is the first time the ejecta have been observed with such high spatial detail and resolution, more sophisticated comparison models are not yet available. We therefore recommend more detailed non-equilibrium ionisation calculations, ideally in 3D, to enable more accurate comparisons with these and future observations of reverse-shocked ejecta in SNRs.

Our observations also reveal spatially resolved, distinct clumps in the ionised limb-brightened ejecta shells of iron and sulphur. We identify five to six compact Fe-rich clumps in [Fe\,\textsc{xiv}] and [Fe\,\textsc{xi}] and four clumps in [Fe\,\textsc{x}] and [Fe\,\textsc{ix}] (with one clump partially obscured by a bright star), which persist across successive ionisation stages and are embedded within a smoother, limb-brightened ejecta envelope. However, [Fe\,\textsc{xv}] at the boundary of the ejecta exhibits less distinct clumps and increased fragmentation. This morphological change is indicative of shock-driven hydrodynamical perturbations that originate as RT or RM instabilities and evolve into KH instabilities at the contact discontinuity, leading to partial fragmentation of clumps in the outer layers while compact cores survive in the inner layers. A direct  comparison with a $\sim$500 yr snapshot of the D6 model shows good qualitative agreement: both the data and the model display similar clump structures in the limb-brightened shells of [Fe\,\textsc{ix}] through [Fe\,\textsc{xiv}], along with comparable fragmentation in [Fe\,\textsc{xv}] (see Figure~\ref{fig:modelclump} and Figure~\ref{fig:modelcontour}). The sulphur shell appears at the ejecta boundary, overlapping with [Fe\,\textsc{xv}] and is therefore expected to experience similar fragmentation, as seen in the D6 model. However, the observed [S\,\textsc{xii}] emission in SNR 0509, although located in a similar region, exhibits compact and distinct clumps rather than a fragmented shell \citep[also shown in][]{Mandal2026}. We also find that instabilities are strongly influenced by shock–ejecta interaction, as illustrated in Figure~\ref{fig:blob}, which shows the morphological evolution of an individual clump under the influence of the reverse shock.

In addition to spatially resolved measurements of ejecta shell extents and clump morphology, our analysis of emission-line profiles from the reverse-shocked ejecta probes the kinematics of the reverse shock. The velocity widths of the lines depend on both thermal and turbulent motions of the shocked gas \citep{Vink2012}. We find a systematic variation of line widths with ionisation state in the eastern limb-brightened shell. Figure~\ref{fig:east_trend} (left) shows an increasing trend in velocity width from the outer high-ionisation shell to the more central, lower-ionisation shell.
Limb-brightened shells let us spatially resolve the successive ionisation zones along the radial direction however the curvature does add a geometric term to the velocity width.
The observed trend suggests an increasing reverse shock speed as it propagates inward, consistent with predictions by \citet{Truelove1999} and \citet{chevalier1982self}. In the western region, the velocity widths also show a general increase from [Fe\,\textsc{xv}] to [Fe\,\textsc{ix}], although the trend is less definitive, with [Fe\,\textsc{x}] exhibiting a similar velocity width as [Fe\,\textsc{xi}] and [Fe\,\textsc{xiv}] within uncertainties . This discrepancy may arise due to the ejecta interacting with a denser ambient medium on the western side. 

Overall, we propose that deep optical observations of spatially resolved ejecta and their distribution provide a powerful new diagnostic for constraining explosion models for SNe Ia and shock–ejecta interactions in Type Ia supernova remnants. The observed clump morphology may additionally reflect the combined effects of hydrodynamical instabilities and a possible Ni-bubble-induced reduction in central ejecta density. This interpretation remains tentative given current uncertainties in ion–ion interactions and non-equilibrium ionisation modelling. Further forward modelling of supernova remnants across a broader range of explosion scenarios and circumstellar environments, together with deep observations of additional Type Ia remnants, will be essential for establishing a statistical understanding of optical ejecta properties and reproducing the observed ionisation structure and line strengths.

\begin{acknowledgements}
JML was supported by basic research funds of the Office of Naval Research.
\end{acknowledgements}

\begin{table*}
\centering
\small
\caption{Wavelength data used to visualise the ejecta clumps in Figure~\ref{fig:iron_clumps}.}
\label{tab:table1a}
\begin{tabular}{lcccc}
   \hline\hline
    Ion & $\lambda_1$ (\text{\AA}) & $\lambda_2$ (\text{\AA}) & $\Delta \lambda_{*1}$ (\text{\AA}) & $\Delta \lambda_{*2}$ (\text{\AA}) \\
    \hline
        {[Fe {\sc xv}]}  & 7001.2 & 7062.45  & 6948.70--6772.45 & 7209.95--7088.70 \\
        {[Fe {\sc xiv}]}  & 5261.2 & 5332.45  & 5191.2--5062.45  & 5568.7--5399.95  \\
        {[Fe {\sc xi}]}  & 7892.45 & 7947.45  & 7763.7--7816.2   & 8038.70--8131.20 \\
        {[Fe {\sc x}]}   & 6381.2  & 6404.95  & 6197.45--6311.20 & 6453.70--6469.95 \\
        {[Fe {\sc ix}]}  & 8194.95 & 8224.95  & 8126.20--8138.70 & 8334.95--8521.20 \\
        {[S {\sc xii}]}  & 7637.45 & 7708.7   & 7306.2--7549.95  & 7783.7--8002.45  \\
        {[Ca {\sc xv}]}  & 5627.45 & 5698.7   & 5517.45--5569.95 & 5756.20--5791.20 \\
   \hline\hline
\end{tabular}
\end{table*}

\begin{table*}
\centering
\small
\caption{Surface brightness contour levels ($\mathrm{erg\,s^{-1}\,cm^{-2}\,arcsec^{-2}}$) used to visualise the ejecta clumps in Figure~\ref{fig:iron_clumps}.}
\label{tab:table1b}
\begin{tabular}{ll}
   \hline\hline
    Ion & SB levels ($\mathrm{erg\,s^{-1}\,cm^{-2}\,arcsec^{-2}}$) \\
    \hline
        {[Fe {\sc xv}]}  & $7.65\times10^{-19}$, $1.53\times10^{-18}$, $3.06\times10^{-18}$, $4.59\times10^{-18}$, $5.36\times10^{-18}$, $6.13\times10^{-18}$ \\
        {[Fe {\sc xiv}]}  & $1.78\times10^{-18}$, $8.90\times10^{-18}$, $1.60\times10^{-17}$, $3.47\times10^{-17}$, $4.45\times10^{-17}$, $4.81\times10^{-17}$, $5.52\times10^{-17}$, $6.06\times10^{-17}$, $6.77\times10^{-17}$ \\
        {[Fe {\sc xi}]}  & $1.25\times10^{-18}$, $2.50\times10^{-18}$, $3.13\times10^{-18}$, $3.75\times10^{-18}$, $4.25\times10^{-18}$, $4.50\times10^{-18}$, $4.75\times10^{-18}$ \\
        {[Fe {\sc x}]}   & $1.19\times10^{-18}$, $2.38\times10^{-18}$, $2.97\times10^{-18}$, $3.86\times10^{-18}$, $4.16\times10^{-18}$, $4.75\times10^{-18}$, $4.87\times10^{-18}$ \\
        {[Fe {\sc ix}]}  & $2.25\times10^{-18}$, $4.5\times10^{-18}$, $6.0\times10^{-18}$, $7.5\times10^{-18}$, $8.25\times10^{-18}$, $9.0\times10^{-18}$, $9.37\times10^{-18}$, $1.01\times10^{-17}$, $1.05\times10^{-17}$ \\
        {[S {\sc xii}]}  & $6.23\times10^{-18}$, $1.43\times10^{-17}$, $1.78\times10^{-17}$, $1.96\times10^{-17}$, $2.14\times10^{-17}$, $2.36\times10^{-17}$, $2.49\times10^{-17}$, $2.67\times10^{-17}$ \\
        {[Ca {\sc xv}]}  & $1.78\times10^{-18}$, $2.67\times10^{-18}$, $3.92\times10^{-18}$, $5.34\times10^{-18}$, $6.23\times10^{-18}$, $7.13\times10^{-18}$ \\
   \hline\hline
\end{tabular}
\end{table*}

\bibliographystyle{mnras}
\bibliography{bibliography}
\end{document}